\documentclass[aps,pre,amsfonts,floatfix, twocolumn,showkeys]{revtex4-1}
\usepackage{amssymb}
\usepackage{amsmath}
\usepackage{amsfonts}
\usepackage{hyperref}
\usepackage{color}
\usepackage{graphicx}
\usepackage{dcolumn}
\usepackage{color}
\usepackage{enumerate}
\definecolor{darkgreen}{rgb}{0.1,0.6,0.3}
\definecolor{gray}{rgb}{0.8,0.8,0.8}
\newcommand{\highlight}[1]{\colorbox{gray}{$\displaystyle #1$}}
\hypersetup{
    colorlinks=true,       
    linkcolor=blue,          
    citecolor=darkgreen,        
    filecolor=magenta,      
    urlcolor= black           
}

\begin{document}

\title{
Evolutionary multiplayer games
}

\author{Chaitanya S. Gokhale$^{1,2}$}
\author{Arne Traulsen$^1$}%
\email{traulsen@evolbio.mpg.de}

\affiliation{%
$^1$Evolutionary Theory Group, Max Planck Institute for Evolutionary Biology, 
August-Thienemannstra\ss e 2 24306 Pl\"{o}n, Germany}%

\affiliation{%
$^2$Present address of Chaitanya S. Gokhale: New Zealand Institute for Advanced Study, \\
Massey University, Auckland, New Zealand}%


\begin{abstract}
Evolutionary game theory has become one of the most diverse and far reaching theories in biology. 
Applications of this theory range from cell dynamics to social evolution. 
However, many applications make it clear that inherent non-linearities of natural systems need to be taken into account.
One way of introducing such non-linearities into evolutionary games is by the inclusion of multiple players. 
An example is of social dilemmas, where group benefits could e.g.\ increase less than linear with the number of cooperators. 
Such multiplayer games can be introduced in all the fields where evolutionary game theory is already well established. 
However, the inclusion of non-linearities can help to advance the analysis of systems which are known to be complex, e.g. in the case of non-Mendelian inheritance. 
We review the diachronic theory and applications of multiplayer evolutionary games and present the current state of the field.
Our aim is a summary of the theoretical results from well-mixed populations in infinite as well as finite populations. 
We also discuss examples from three fields where the theory has been successfully applied, ecology, social sciences and population genetics.
In closing, we probe certain future directions which can be explored using the complexity of multiplayer games while preserving the promise of simplicity of evolutionary games.
\end{abstract}


\maketitle


\section*{Introduction}
\label{intro}

Game theoretic reasoning can be traced back to the Babylonian Talmud \citep{aumann:PNAS:2006}, 
but possibly the first mathematical proof using game theory was about the game of chess by Zermelo \citep{zermelo:PCM:1913,schwalbe:GEB:2001}.
Typically, the initiation of evolutionary game theory is ascribed to Morgernstern and von Neumann, who published the first seminal treatise on game theory \citep{neumann:book:1944}. 
While most of the theory developed therein is for two-player games, as Nash pointed out \citep{nash:PNAS:1950}, it indeed has a section on multiplayer games.
However, the multiplayer games tackled by Morgernstern and von Neumann were the so called \textit{cooperative} games where the interacting players can form coalitions.
After developing a theory for non-cooperative games, where the individuals are driven by purely selfish motives and no sense of collaboration, Nash promptly applied it to another famous game, poker \citep{nash:AM:1951}.

The formal use of game theory in biology can be ascribed to Fisher, who used indirect game theoretic reasoning to 
tackle the question of why sex ratio in mammals usually tends to $1:1$ \citep{fisher:book:1930}. 
Mathematical arguments of a similar form were already presented in the nineteenth century \citep{darwin:book:1859,dusing:JZN:1884,edwards:TPB:2000,osborne:WA:1996}.
For the use of game theory, Fisher was later explicit in stating \citep{fisher:JE:1958} `The relation between species, or among the whole assemblage of an ecology, may be immensely complex; and at Dr. Cavalli's invitation I propose to suggest that one way of making this intricate system intelligible to the human mind is by the analogy of games of skill, or to speak somewhat more pretentiously, of the Theory of Games.'
\citet{lewontin:JTB:1961} gave a more complete introduction of game theory to biologists, while
\citet{maynard-smith:Nature:1973} formally presented what we know of today as evolutionary game theory.

Although classical game theory originates from economics \citep{neumann:book:1944,nash:AM:1951}, evolutionary game theory forgoes a typical assumption of classical game theory: rationality. 
In evolutionary game theory, natural selection is the dominant force. Individuals are born with fixed strategies. They interact with each other and receive payoffs according to a payoff matrix based on their strategies. 
Strategies which receive the higher payoff are said to be more successful than those which do not. 
These successful strategies spread in the population at the cost of other less successful strategies. 
Understanding this dynamical process is the mainstay of evolutionary game dynamics \citep{sandholm:book:2010}.

The initial focus of evolutionary game theory was on the concept of evolutionarily stable strategies \citep{maynard-smith:Nature:1973}.
Evolutionary stability is a refinement of the concept of a Nash equilibrium.
This leads to ideas such as an `unbeatable strategy' \citep{hamilton:Science:1967} or an `evolutionarily stable strategy' \citep{maynard-smith:book:1982}.
A strategy is defined to be `unbeatable' or an ESS if a small number of individuals playing a different strategy cannot invade a population playing it.
While `unbeatable' strategies dominate all the invading mutants, a weak ESS 
can allow the invasion of initially neutral mutants.
The notion of evolutionarily stable strategies was already generalised to multiplayer games in the early eighties \citep{palm:JMB:1984}.

Knowing if a strategy is an ESS is very useful, but an important question is if such a strategy is attainable. 
The work of Taylor, Jonker and Zeeman \citep{taylor:MB:1978,zeeman:LN:1980} extended the realm of evolutionary game theory to include dynamics, which led to a straightforward relation to the ESS concept \citep{hofbauer:book:1998}.

Key advances were also made when spatial structure \citep{nowak:Nature:1992b} and finite populations \citep{thomas:Biosystems:1981,ficici:proceedings:2000,nowak:Nature:2004} were included in evolutionary games. 
Addressing spatial structure in evolutionary games sparked a whole field of its own and would need to be reviewed separately -- we thus do not consider spatial structure here.  

Going from infinite populations to finite ones, the questions changed from evolutionary stability in the deterministic regime to the properties pertaining to fixation, extinction, maintenance of multiple strategies, etc.
The analysis, however, was mostly limited to two-player games \citep{nowak:book:2006,miekisz:LN:2008}.
We highlight recent results obtained in the field of evolutionary multiplayer games both for infinitely large populations as well as finite population.

\section*{From pairwise contests to social interactions}

The leap from chess to poker was crucial in the development of the theory of games.
Economics continued with multiplayer analysis \citep{kim:GEB:1996,wooders:GEB:2006,ganzfried:IJCAI:2009} but a similar growth pattern was not reflected in evolutionary game theory. 
Instead, the simplicity of evolutionary games helped spread its applications from genes and cells \citep{bach:EJC:2001,axelrod:PNAS:2006,basanta:bookchapter:2008}, between individuals or communities \citep{axelrod:book:1984,turner:Nature:1999,archetti:JTB:2000,turner:AmNat:2003,frey:PA:2010} and even across species \citep{poulin:JTB:1995,vickery:EEc:2010}.

However, even complex biological phenomena can be incorporated in evolutionary games without compromising the simplicity of the theory.
The past decade saw an explosive growth in the inclusion of finite population dynamics in evolutionary games \citep{thomas:Biosystems:1981,nowak:Nature:2004,nowak:book:2006,traulsen:bookchapter:2009}.
This has substantially increased the scope of the theory, while leading to beautiful and simple results
in its own right \citep{nowak:Nature:2004,lessard:JMB:2007}. 
Similarly, the consideration of non-linear interactions in multiplayer games could open new avenues of research. 
Herein, we list the current state-of-the-art in evolutionary game theory dealing with non-linear interactions brought about by multiple players.
Such inclusions have the potential to demonstrate novel dynamics which is not possible in conventional two-player games \citep{broom:BMB:1997,bukowski:IJGT:2004,gokhale:JTB:2011}.

\subsection*{Replicator dynamics}

Traditionally, 
`Evolutionary game theory, [. . .], describes evolution in phenotype space' \citep{nowak:Science:2004}. 
The different phenotypic traits are termed strategies. 
At the core of evolutionary game dynamics lies the replicator equation.
It was named so after taking inspiration from the concept of `replicators' from Dawkins \citep{dawkins:book:1982,schuster:JTB:1983}.
For a detailed connection between the replicator equation and other fundamental dynamic equations such as the quasi-species equation and the replicator--mutator equation, see \citep{page:JTB:2002}.

The replicator equation allows the frequencies of the different types in the population to determine the fitness landscape rather than setting the fitness of each type to be constant (constant fitness being a special case of the replicator dynamics).
Taking a bottom-up approach to the replicator equation, consider two types in an infinitely large population, $A$ and $B$. 
The frequencies of the two types are given by $x$ and $1-x$. 
The interactions between the two types can be expressed by a matrix such as
\begin{eqnarray}
\bordermatrix{ & \text{A} & \text{B} \cr
\text{A} & a_1 & a_0 \cr 
\text{B} & b_1 & b_0
\cr}, 
\label{2plmat}
\end{eqnarray}
This payoff matrix shows that when an $A$ individual interacts with another $A$ individual it gets $a_1$ and when interacting with a $B$ individual it gets $a_0$.
From this payoff matrix, we can calculate the average payoff of both the strategies,
$\pi_A = a_1 x + a_0 (1-x)$ and $\pi_B = b_1 x + b_0 (1-x)$.

\begin{figure*}
\begin{center}
\includegraphics[scale=0.5]{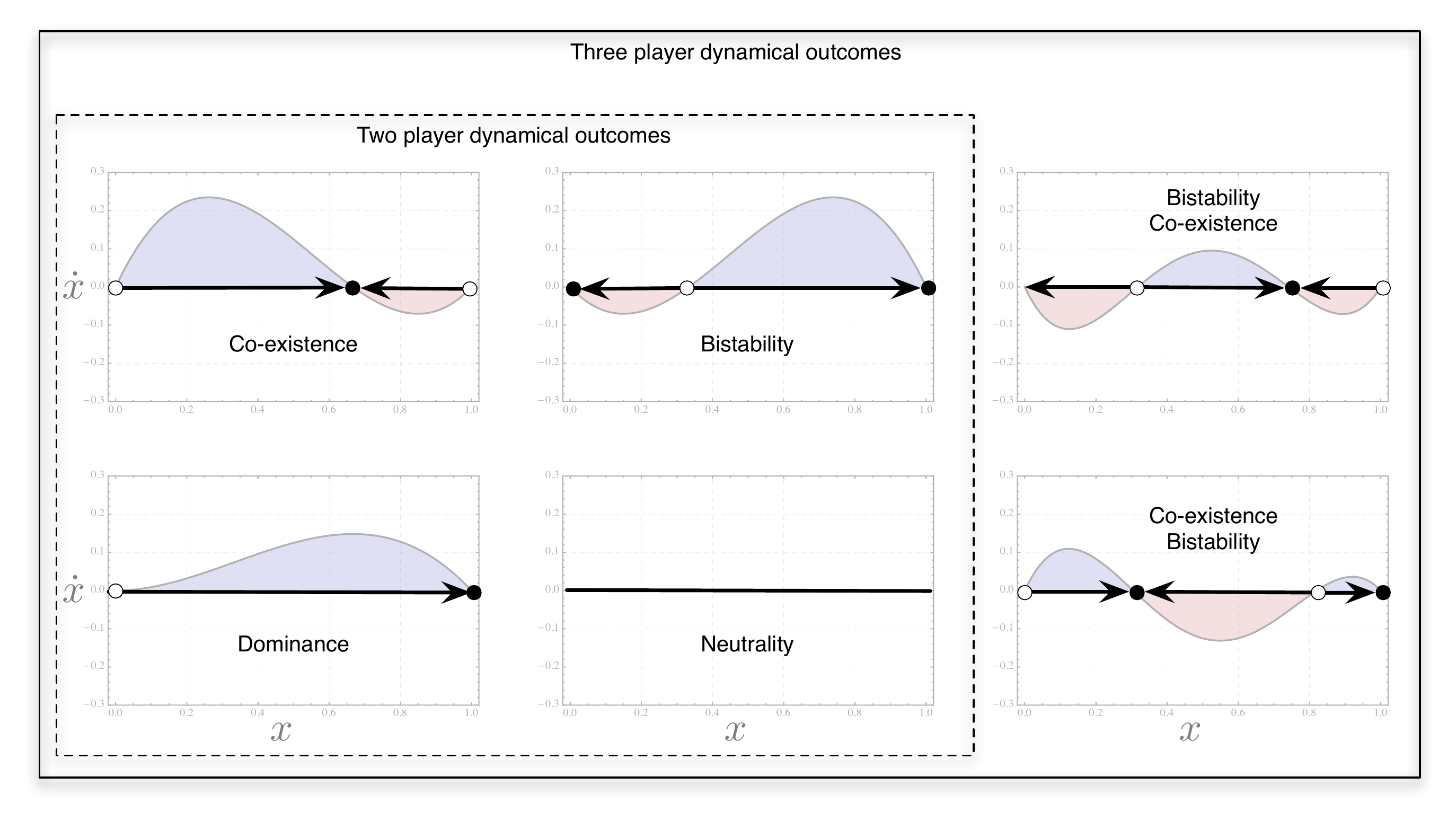}
\caption{
\label{outcomes}
Possible outcomes in a two-player game and comparatively in a three-player game with two strategies $A$ and $B$.
Since the 
selection gradient $f_A - f_B$ for a two-player game is linear, the possible outcomes can include at most one internal equilibrium point, which can be either stable or unstable. 
Increasing the number of players increases the complexity of the dynamical equation by making the selection gradient non-linear.
For a three-player case the equation is a quadratic polynomial and can hence contain at most two internal equilibrium points, which can be alternatively stable and unstable.
}
\end{center}
\end{figure*}

We can interpret these average payoffs as fitnesses of the two strategies directly denoted by $f_A = \pi_A$ and $f_B = \pi_B$. 
How this fitness actually relates to the payoff is a question pertaining to the particular context of the model we are studying,
in particular, in finite population this issue can be of importance \citep{wu:PRE:2010}. 
For the time being, we assume them to be the same.
Again, following classical selection ideas if this fitness is greater than the average fitness of the population, then the frequency of that type increases over time and vice versa.
These concepts can be formally written down in the form of a differential equation which tracks the change in $x$ over time,
\begin{eqnarray}
\dot{x} &=& x (1-x) (f_A - f_B) \nonumber \\
&=& x (1-x) ( (a_1-a_0-b_1+b_0) x +a_0-b_0)
\end{eqnarray}
Thus, the evolutionary game is introduced in the dynamics via the fitness of the strategies.
There are three possible solutions to this equation, strategy $A$ goes extinct, $x = 0$ or the whole population consists of $A$ players, $x = 1$ and lastly when the two strategies have equal fitness, $f_A = f_B$ \citep{bishop:AAP:1976} which is when 
\begin{eqnarray}
\label{eq:internal}
x^\ast = \frac{b_0-a_0}{a_1-a_0-b_1+b_0}.
\end{eqnarray}
See Fig.~\ref{outcomes} for a summary of the possible dynamics.

\subsubsection{Equilibrium points in multiplayer games}

For general multiplayer games, the story is a bit more complicated, and hence let us begin with the simplest case of three players, still with the above two strategies $A$ and $B$.
As before, the frequencies are given by $x$ and $1-x$ but now the interactions are given by the following payoff matrix:
\begin{eqnarray}
\bordermatrix{ & \text{AA} & \text{AB}	& \text{BB} \cr
\text{A} & a_2 & a_1 & a_0 \cr 
\text{B} & b_2 & b_1 & b_0 \cr}, 
\label{3plmat}
\end{eqnarray}
where the focal individual is denoted by the row player.
Since it is a three-player game, there are two other players, who are denoted as the column players.
They can be either $AA$, $AB$, $BA$ or $BB$.
Herein, we assume that playing with an $AB$ is the same as playing with $BA$.
Thus, the labels $A$ and $B$ matter, but not their ordering.
If the ordering of players would matter, then we would have two different payoffs entries for $AB$ and $BA$.
In this simple case, increasing the number of players increases the number of payoff entries.
The average payoffs of the two strategies are a polynomial function of the frequency, 
$\pi_A = a_2 x^2 + 2 a_1 x (1-x) + a_0 (1-x)^2$ 
and 
$\pi_B = b_2 x^2 + 2 b_1 x (1-x) + b_0 (1-x)^2$.
Again, as before, equating fitness with payoff we can write down the replicator equation, 
\begin{eqnarray}
\dot{x} &=& x (1-x) (f_A - f_B)  \\ \nonumber
&=& x (1-x)  ( x^2 (a_0-2 a_1+a_2-b_0+2
   b_1-b_2) \\ \nonumber 
   && +2 x
   (-a_0+a_1+b_0-b_1)+a_0-b_0 ). 
\end{eqnarray}
Again, there exist two trivial roots, $x=0$ and $x=1$, as previously but now there is a possibility of two more roots to exist, as the payoff difference is polynomial of degree 2 in $x$ (Fig.~\ref{outcomes}).
We can immediately extend this analysis to an arbitrary number of $d$ players, 
where the possible number of roots (except the trivial) can be $d-1$
\citep{hauert:JTB:2006a}.
For a game with two strategies, tracking the frequency of a single strategy provides all the information about the dynamics. 
For a game with $n$ such strategies, we need to know the time evolution of $n-1$ variables. 
Hence, for any $d$ player game with $n$ strategies, the dynamics proceeds on an $n-1$ dimensional simplex.
Since in each dimension the number of equilibria possible in the interior are $d-1$, in all there can be $(d-1)^{n-1}$ distinct internal equilibria at most \citep{gokhale:PNAS:2010,han:TPB:2012}.

In evolutionary game theory, fixed points of the dynamical system are the particular composition of strategy frequencies where all the strategies have the same average fitness. 
Interpreting these fixed points biologically, they predict a co-existence of different types in a population and the maintenance of polymorphism. 
The study of the properties of such equilibrium points has a long-standing history in classical game theory, evolutionary game theory and population genetics 
\citep{karlin:TPB:1972b,bishop:AAP:1976,karlin:JMB:1980,cannings:JTB:1993,broom:PRSA:1993,broom:BMB:1997,altenberg:TPB:2010}.
The number of fixed points is an important property of each concrete system, but what is the generic number of fixed points in a system?
More precisely, what is the probability that we will have a system with a certain number of fixed points?
This can be analysed by an exhaustive study of the maximal number of equilibrium points of a system and the attainability of the patterns of evolutionarily stable strategies in an evolutionary system, both analytically and numerically \citep{karlin:TPB:1970,karlin:JMB:1980,maynard-smith:book:1982,vickers:JTB:1988,cannings:JTB:1988,broom:JMB:1994,ganzfried:IJCAI:2009,han:TPB:2012}.

A method for addressing such general questions is via the study of randomly drawn evolutionary games. 
If the payoff matrices are drawn randomly from an arbitrary distribution then we can ask the question, what are the probabilities of observing a certain number of (stable) equilibria?
For such randomly drawn two-player games with $n$ strategies (such that all payoff entries are drawn from the same distribution), the probability that there exists an isolated internal equilibrium is $2^{1-n}$ and the probability that a given equilibrium is stable is at most $2^{-n}$.
For $n=2$ this is exactly equal to $1/4$.
Extending such analysis to multiplayer games, we see that given an equilibrium for a $d$ player two-strategy game, the probability of it being stable/unstable is just $1/2$.
For $d$ player games with $n$ strategies, the probability of having a given number of equilibria has been calculated explicitly for several given cases, 
but not yet in closed form across the number of players.
For example, the probability that a three-player game with two strategies has two internal isolated equilibria is analytically determined to be $\frac{61}{450} \approx 0.136$ and well corroborated by simulations \citep{han:TPB:2012}.

This analysis in addition to the translation of the game theoretic framework to population genetics helps us draw a number of parallels. 
For example, going back to classical population genetics \citep{rowe:JTB:1987,rowe:JTB:1988,karlin:PNAS:1969,karlin:JMB:1980}, 
a current study on multiplayer games \citep{han:TPB:2012} provides a proof of the conjecture put forth by Karlin and Feldman on the maximum number of fixed points in a deterministic model of viability selection among $n$ different haplotypes. 
While the evolutionary game theory-based proof requires the assumption of random matching, an alternative approach presented by \citet{altenberg:TPB:2010} 
proves the conjecture in all its generality.

\subsection*{From infinite to finite populations}

The replicator dynamics describes the dynamics of strategy frequencies in an infinitely large population.
Clearly, this is an approximation.
It has been well acknowledged that the studies in finite population have the capacity to challenge the results based on the infinite population assumption \citep{fogel:EM:1998,nowak:Science:2004}. 
Early on, using principle from the philosophy of science, \citet{thomas:Biosystems:1981} demonstrated the shortcomings of the classical ESS theory when applied to finite populations.
The classical concept of an ESS is shown to be neither necessary nor sufficient to describe evolutionary stability in small finite populations, while for large populations it is necessary but not sufficient \citep{nowak:Nature:2004}.
Since then, there has been a rapid development in the field of finite population analysis of evolutionary games
\citep{traulsen:PRL:2005,imhof:PNAS:2005,fudenberg:JET:2006,antal:BMB:2006,traulsen:PRE:2006c,traulsen:JTB:2007b,hauert:science:2007,ohtsuki:JTB:2007c,lessard:JMB:2007,kurokawa:PRSB:2009,gokhale:PNAS:2010,imhof:PRSB:2010,wu:PRE:2010,zhou:PRE:2011,wu:JMB:2012,wu:PLoSCB:2013}.
We begin by introducing stochastic dynamics for two-player games and then generalise the results for multiplayer games.

\subsubsection*{Moran process}

The replicator dynamics can be viewed as a limiting deterministic case for a variety of stochastic processes.
For finite populations, we can imagine a number of different microscopic processes for the transmission of strategies.
Choosing the pairwise comparison process as a microscopic process results in the imitation dynamics in the limit of infinitely large population sizes \citep{traulsen:PRL:2005,traulsen:PRE:2012}.
For our purpose, we will focus on a variant of the Moran process.
The classical Moran process proposed in evolutionary theory \citep{moran:book:1962} is a one-step process with a constant population size.
The Moran process in evolutionary game theory \citep{taylor:BMB:2004} usually consists of picking an individual according to its fitness for reproduction and then picking another individual randomly for death.
The reproducing individual begets an offspring with the same strategy as its own, and the individual chosen for death is removed from the population.
As these events take place in the same time step, the population size is conserved.
Fitness determines the probability of an individual to be chosen for reproduction. 
Thus, the transition probabilities of such a process are given by
\begin{eqnarray}
T_j^+ &=& \frac{j f_A}{j f_A + (N-j) f_B} \frac{N-j}{N} \nonumber \\
T_j^- &=& \frac{(N-j) f_B}{j f_A + (N-j) f_B} \frac{j}{N}
\label{eq:transprob}
\end{eqnarray}
while the system remains in the same state with probability $1-T_j^+ - T_j^-$.
The concept of an intensity of selection is by no means new to evolutionary theory.
However, traditionally, the selection coefficient was just the relative difference between the fitness of two types.
In evolutionary game theory, we consider an independent variable which controls the effect of selection.
Following the logic from the previous paragraph, fitness would determine the probability of reproduction.
Hence, the selection term is introduced when calculating the fitness from the payoff.
Our earlier assumption of $f_i = \pi_i$ thus no longer holds, but rather the fitness is a non-decreasing function of the payoff. The importance of the game for fitness is controlled by the selection intensity, which we term as $w$.
We can thus tune the intensity of selection to control the impact of the game on the fitness.
When selection is weak, all strategies have almost the same fitness while for higher intensities of selection, the game matters.

\subsubsection*{Fixation probability}

For games where the mutations rates are very low, an individual with a new strategy can either go extinct or go to fixation \citep{fudenberg:JET:2006,wu:JMB:2012}.
Calculating the probability of fixation of such a new mutant is very useful in studying the dynamics of the spread of strategies.
Hence, in a population of size $N$ consisting of $N-1$ $B$ individuals and a single $A$ individual, the probability $\rho_1^A$ that it will take over the population is given by \citep{karlin:book:1975,nowak:book:2006},
\begin{eqnarray}
\rho_1^A = \frac{1}{1+\sum_{k=1}^{N-1} \prod_{j=1}^k \frac{T_j^-}{T_j^+}}.
\end{eqnarray}
Under neutrality, $T_j^- = T_j^+$, the fixation probability is simply that of a neutral mutant, $1/N$. 
This is a well-known result in classical population genetics for the fixation probability of a neutral allele \citep{goel:book:1974,kimura:book:1983}.

The game enters the dynamics via the inclusion of fitness which in the above case is packaged via $ \tfrac{T_j^-}{T_j^+} =\tfrac{f_B}{f_A} $, as the transition probabilities given in Eqs.~\eqref{eq:transprob}.
Hence, now, it matters how the fitness function maps the payoffs and selection intensity to fitness.
Assuming a linear payoff to fitness mapping, we have $f_S = 1- w + w \pi_S$. 
This leads to the ratio of transition probabilities $ \frac{T_j^-}{T_j^+}= \frac{1-w+w \pi_B}{1-w+w \pi_A}$.
Further, we assume that selection is weak $w \ll 1$.
This simplifies matters, as it corresponds to a linear approximation with the neutral case as a reference.
There are various ways in which we can interpret why selection would be weak in the first place even without an external parameter governing selection intensity \citep{tarnita:PNAS:2011}.
This simplifies $ \frac{T_j^-}{T_j^+}= \frac{1-w+w \pi_B}{1-w+w \pi_A} \approx 1- w (\pi_A -\pi_B)$.
Alternatively, we can define $f_S = \exp[w \pi_S]$, in which case without the assumption of weak selection we have a simpler form of $ \frac{T_j^-}{T_j^+} = e^{- w (\pi_A -\pi_B)}$.
From these approaches, we see that the quantity of interest is the difference in the payoffs, which actually forms a condition for the derivation of further results.

With this setup now we can ask the traditional question of evolutionary stability but in finite populations.
The use of fixation probability to characterise evolutionary stability was first implemented in \citep{nowak:Nature:2004}.
Two questions are of key importance in determining the potential of a strategy to take over the populations:
\begin{itemize}
\item[(i)] When is the fixation probability of a strategy greater than neutral (One-third rule)?
\item [(ii)] When is the fixation probability of a strategy greater than the fixation probability of the other strategy (Risk Dominance)?
\end{itemize}

Tackling the first point for two-player games leads to a beautiful inequality popularly known as the one-third rule,
\begin{eqnarray}
\rho_1^A > 1/N \text{ if } x^\ast < 1/3,
\end{eqnarray}
which holds for a 
wide range of evolutionary processes \citep{nowak:Nature:2004,imhof:jmb:2006,traulsen:PRE:2006c,lessard:JMB:2007}.
In words, this means that assuming a large finite population with random matching, selection favours strategy $A$ replacing the resident strategy $B$ if the internal equilibrium $x^\ast$ (as defined in Eq.~\ref{eq:internal}) is less than one-third.

Tackling the second point for two-player games, strategy $A$ will replace strategy $B$ with a higher probability, that is $\rho_1^A > \rho_1^B$ than vice versa if $N a_0 + (N-2) a_1 > (N-2) b_0 + N b_1$.
For large enough populations, this is simply $a_0 + a_1 > b_0 + b_1$.
Analogous results to this condition exist in a deterministic setting as well which were discussed in \citep{kandori:ECO:1993}.
This condition is also holds true for the Moran process with a variety of birth--death processes under weak selection and for some special processes for any intensity of selection \citep{nowak:Nature:2004,antal:JTB:2009a}.

Performing a similar analysis for multiplayer games, albeit via more complicated mathematics, yields the generalisation of the so-called one-third rule and the risk dominance condition \citep{kurokawa:PRSB:2009,gokhale:PNAS:2010}.
For large but finite populations,
\begin{itemize}
\item[(i)] the fixation probability of a strategy is greater than neutral for a $d$ player game, $\rho_1^A > 1/N$ 
if the condition $ \sum_{k=0}^{d-1} (d-k)a_k > \sum_{k=0}^{d-1} (d-k)b_k
$ is fulfilled, and
\item[(ii)] the fixation probability of $A$ is large than the one of $B$, $\rho_1^A > \rho_1^B$, if 
$\sum_{k=0}^{d-1} a_k > \sum_{k=0}^{d-1} b_k$.
\end{itemize}
The condition for overcoming neutrality is now no longer connected to the equilibrium point explicitly as was in the two-player case making the one-third rule special.
Furthermore, this result is valid for all processes within the domain of Kingman's coalescence \citep{lessard:DGGA:2011b}.
While the one-third rule is derived under weak selection, for strong selection the evolutionary stability criterion for finite populations becomes equivalent to the evolutionary stability condition in infinite populations \citep{hofbauer:book:1998,nowak:book:2006,traulsen:PRE:2006c}. 
For intermediate selection intensities, the fixation probability can have at most a single maximum or minimum
\citep{wu:Games:2013}.

\subsubsection*{Fixation time}

Another interesting property of the dynamics in finite populations is the average time to fixation. 
Specifically, the times of interest are 
(i) the unconditional fixation time (the average time required to reach any of the boundaries, all $A$ or all $B$) 
and
(ii) the conditional fixation time (the average time required to reach a given boundary given that it is reached).
Expressions for these times are available from standard textbooks on stochastic processes \citep{karlin:book:1975,ewens:book:1979}.
Quite often, we are interested in fixing a strategy in a population.
The expression for the conditional fixation time is then useful as it is the time required for a mutant to reach fixation given that it does reach fixation.
Starting with a single $A$ individual, we have the conditional fixation time
\begin{eqnarray}
\tau_1^A = \sum_{k=1}^{N-1} \sum_{l}^{k} \frac{\rho_l^A}{T_l^+} \prod_{j=l+1}^{k}\frac{T_j^-}{T_j^+}.
\end{eqnarray}
As described above, the game is introduced via $T_j^-$ and $T_j^+$ (that also enter in the fixation probability $\rho_l^A$) which is a ratio of transition probabilities which in turn depend on the payoff-determined fitnesses of the strategies.
Under weak selection, the form of both the unconditional and conditional fixation time has been derived for a variety of processes \citep{altrock:NJP:2009,wu:PRE:2010}.
In particular for the conditional fixation time the first-order series expansion is of the form,
\begin{eqnarray}
\label{eq:CondFixTimeFOE}
\tau^{A}_{1} \approx [\tau^{A}_{1}]_{w=0} + w \left[\frac{\partial}{\partial w}\tau^{A}_{1}\right]_{w=0}.
\end{eqnarray}
The zeroth-order term is $N (N-1)$ while the first-order term was explicitly calculated for two-player games e.g.\ in \citep{altrock:NJP:2009}.
Interestingly, for a strategy which always has a fitness advantage that varies with the frequency, 
the first-order term can be positive.
This means that even though a strategy has a fitness advantage it takes longer to fix than a neutral mutant.
This seemingly counterintuitive case is analysed in detail in \cite{altrock:PRE:2010,altrock:JTB:2012}.
Performing a similar analysis for multiplayer games involves first calculating the first-order term.
For $d$ player games, the first-order correction can be exactly calculated for a given process
or in more general terms \citep{wu:PRE:2010}.
For example, for the Moran process with an exponential payoff to fitness mapping, $f_S = \exp[w \pi_S]$, we have
\begin{eqnarray}
\left[\frac{\partial}{\partial w}\tau_{1}^A\right]_{w=0}
&=& N \sum_{i=1}^{N-1} \Delta \pi (i) [ i (H_{N-1} - H_{i-1} - H_{N-i}) \nonumber \\
&&+ N H_{N-i}] - \Gamma (2 + N H_{N-1})
\end{eqnarray}
where $\Delta \pi (i) = \pi_A (i) - \pi_B (i)$, $H_{i} = \sum_{k=1}^{i}1/k$ is the harmonic number 
and $\Gamma$ is a function of the payoff parameters, see Eq.~(S13) in \citep{gokhale:PNAS:2010}.
Analysing this effect in multiplayer games, we see that increasing the number of players 
can intensify the effect of this so-called stochastic slowdown \citep{wu:Games:2013}.

\subsubsection*{Mutation selection equilibrium}

For intermediate mutation rates \citep{wu:JMB:2012}, a new mutation can occur even when the first mutation is at an intermediate frequency.
This results in a polymorphic population, and the concept of fixation is not very useful anymore for characterising the system.
In this case, the success of a strategy can be characterised by the average frequency of a strategy in the mutation selection equilibrium or in short, the average abundance.
By success we mean that in the average abundance of a strategy is greater than that of the other.
An alternative approach based on a limiting case of vanishing noise \citep{foster:TPB:1990} was employed for the the stochastic stability of three-player games in \citep{kaminski:BMB:2005}. This approach can also be extended to spatial games \citep{miekisz:JSP:2004,miekisz:PA:2004}.

The average abundance of a strategy for arbitrary mutation rates and weak selection was derived explicitly by \cite{antal:JTB:2009a}.
An extension of the same model for multiple strategies followed \citep{antal:JTB:2009b}.
The derivation of the result depends heavily on neutral coalescence theory \citep{kingman:JAP:1982,wakeley:book:2008}.
Hence, the results can only be viewed as weak selection approximations.
Extending the analysis to multiplayer games, it is possible to write down the average abundance of a strategy for a game with e.g. three players \citep{gokhale:JTB:2011}, a closed form for an arbitrary number of player games has not been obtained yet.

For two-player multiple-strategy games (and population structure as well), transforming the results for the average abundance into a slightly different form, it is possible to condense all the information about the process and the population structure under weak selection into a single variable, the so-called structure parameter $\sigma$ \citep{tarnita:JTB:2009}. Then, the condition for $A$ to be more abundant than $B$ is $\sigma a_1 + a_0 > b_1 + \sigma b_0$. 
In case of well-mixed populations we have $\sigma = (N-2)/N$. As $N \rightarrow \infty $ we have $\sigma = 1$ recovering the standard condition for risk dominance. 
Different $\sigma$-values have been calculated for a variety of evolutionary games with two strategies in differently structured populations in \cite{tarnita:JTB:2009}.
The use of such a $\sigma$ parameter results in a single inequality capable of determining whether a strategy had a higher average abundance in the mutation selection equilibrium, a result that can further be extended to multiple strategies in structured populations \citep{tarnita:PNAS:2011}.
For $d$ player games, it can be shown that instead of a single $\sigma$ parameter we would require $d-1$ such $\sigma$ parameters to capture the effects of the process \citep{wu:Games:2013}.
For example, the condition that strategy $A$ is more abundant than $B$ requires
\begin{eqnarray}
\sum_{\substack{0\le i\le d-1\\ i\neq i^*}}\sigma_ia_i+a_{i^*}>\sum_{\substack{0\le i\le d-1\\ i\neq i^*}}\sigma_ib_{d-1-i}+b_{d-1-i^*}.
\end{eqnarray}
For game dynamics in a large well-mixed population when mutations are negligible, the condition for higher abundance reduces to risk dominance.

\section*{Application of multiplayer games}

Due to its generality, the replicator equation encompasses a variety of biological contexts from ecology to population genetics and from prebiotic to social evolution \citep{schuster:JTB:1983} and hence became a popular tool amongst behavioural ecologists, population geneticists, sociologists, philosophers and also back among economists \citep{samuelson:AER:1985,damme:EEcR:1994,sandholm:book:2010}.
This similarity sometimes allows us to transfer results from one field to another \citep{traulsen:JTB:2012,han:TPB:2012}.
Probably, one of the first applications of multiplayer game theory was to the war of attrition provided by \cite{haigh:AAM:1989}.
In the following section, we will see how not just evolutionary game theory in general, but also multiplayer evolutionary games in particular pervade a plethora of disciplines from ecology to social sciences.

\subsection*{Ecology}
From a dynamical systems point of view, the continuous time frequency-dependent selection equations from population genetics, the Lotka-Volterra equations from ecology and the replicator equations all are closely related \citep{riechert:ARES:1983,sigmund:MB:1987,cressman:JTB:1988,hofbauer:book:1998,page:JTB:2002,cressman:book:2003}. 
The set of replicator equations for $n$ strategies is mathematically equivalent to the well-known Lotka-Volterra equations for $n-1$ species in ecology \citep{hofbauer:book:1998}. 
The dynamical equations developed by Lotka and Volterra pre-date the replicator equation by almost half a century \citep{volterra:JCIEM:1928,lotka:JWAS:1932}. 
Hence, in a sense, `Ecology is the godfather of evolutionary game theory' \citep{hofbauer:book:1998}.

An important development in the theoretical understanding of multiple interactions in animals was pushed forth by the introduction of biological markets 
\citep{noe:Ethology:1991,noe:TREE:1995,noe:book:2001}.
Empirical evidence from the wild about coalition formation and multi-party interactions is available for almost a century now ranging from cormorants \citep{bartholomewJr:Condor:1942} to killer whales \citep{smith:CJZ:1981}.

A long-standing question in a complex ecology is the evolution and maintenance of biodiversity \citep{may:Nature:1972}. 
If nature is red in tooth and claw as often envisioned through natural selection, then the persistence of many different species together requires an explanation. 
One way in which multiple species and their interactions have been analysed is via multiplayer games \citep{damore:Evolution:2011}. 
However, instead of calling these interactions as multiplayer, one should consider them as interspecies interactions \citep{schuster:BC:1981} (although they can of course be understood using the same dynamical equations as for multiplayer games).  
If more than two individuals of the same species interact with each other, then we call this a multiplayer game. 
However, it is of course possible that multiple players of one species interact with multiple players of another species.
For example, we can have interactions between many cleaner wrasses and their client. 
For each cleaner fish, it interacts with one client fish at a time and hence we can consider this as a two-player game. 
In this case, the interactions within species are ignored \citep{schuster:BC:1981}.
However, this simplifying assumption can be relaxed \citep{schuster:BC:1981c} leading to further complications. 
As an example, we discuss simple mutualistic interactions between two species (excluding self interactions) where more than two individuals are interacting. 

Consider two species $A$ and $B$. 
It was shown previously \citep{bergstrom:PNAS:2003} that if the two species are locked in a mutualistic relationship with each other, then the species which evolves slower can get a larger share of the benefit. 
The interaction occurs between two different species. 
In each species, we have two different types of individuals, the generous ones $G$ and the selfish $S$. 
The generous individuals contribute to the mutualistic endeavor, while the selfish withhold some contribution. 
Since the interaction is assumed to be mutually beneficial, both species being selfish is not a viable option. 
However if one of the species evolves slower than the other, the slower evolving species can get away with being selfish while forcing the other to be generous. 
This shares some similarity with extortioner strategies recently found in repeated two-player games \citep{press:PNAS:2012,hilbe:PNAS:2013}.
The effect was termed as the Red King effect as opposed to the Red Queen from classical ecology where there is pressure on a species to evolve faster. 

While this analysis was based on two-player games, mutualism can also be explored using multi-partner interactions \citep{noe:book:2001,archetti:EL:2011}.
Multiplayer games make the dynamics and in turn their solutions non-linear. 
This changes the size of the basins of attraction of the equilibria in which one of the species is selfish while the other is forced to be generous. 
Ultimately, this can reverse the Red King effect \citep{gokhale:PRSB:2012}. 

Furthermore addition of multiple players provides a way of extending the analysis in a variety of ways. 
Threshold effects (where a certain number of cooperators are necessary to generate a public good, see e.g. \citep{abouchakra:PLOSCB:2012,pena:JTB:2014}), asymmetric number of players and their interactions with asymmetric growth rates can be explored which reveal a rich dynamics that is possible in mutualisms.

\subsection*{Population genetics}

While ecology was a natural playground for playing evolutionary games, population genetics was not far behind.
Often, evolutionary games are described as a theoretical framework for describing dynamics at the phenotypic level \citep{nowak:Science:2004}, but the similarity of the dynamical equations used with those of population genetics did not go unnoticed.

Early on, the comparisons between game theoretic reasoning and standard population genetic models were explored.
For example, the famous example of Fisher on sex ratios was revisited in \citep{sigmund:TPB:1987} and the fundamental theorem of natural selection in \citep{sigmund:MB:1987}.
Even diploidy was incorporated using multiplayer evolutionary games \citep{rowe:JTB:1987,rowe:JTB:1988} although they were not named so.
In general, we can thus approach population genetics with two different views, with either the game dynamics given by the gene dynamics or as a dynamics on the phenotypic level which occurs based on a known genetic setup.

Biological interactions can be highly non-linear \citep{shirakihara:RPE:1978}.
Especially the non-linear epistatic nature of genetic interactions is a recent subject of interest.
Thinking of strategies as alleles can help apply some of the results of game theory directly to population genetics.
This approach usually restricts the analysis to haploid populations. 
Yet, recent evolutionary game theory has been successful in deriving results for the equilibrium points \citep{han:TPB:2012} and fixation probabilities and fixation times in diploids \citep{hashimoto:JTB:2009}.
The use of multiplayer evolutionary game theory as was employed by \citet{rowe:JTB:1988} attributes each genotype a different strategy. 
Two-player evolutionary games can be used to address population genetic effects of drive elements \citep{traulsen:JTB:2012}. 
We provide an example for an application of multiplayer games in population genetics which is able to handle non-linearities and non-Mendelian inheritance patterns, for example, the dynamics of the Medea allele.

\begin{figure}
\begin{center}
\includegraphics[scale=0.2]{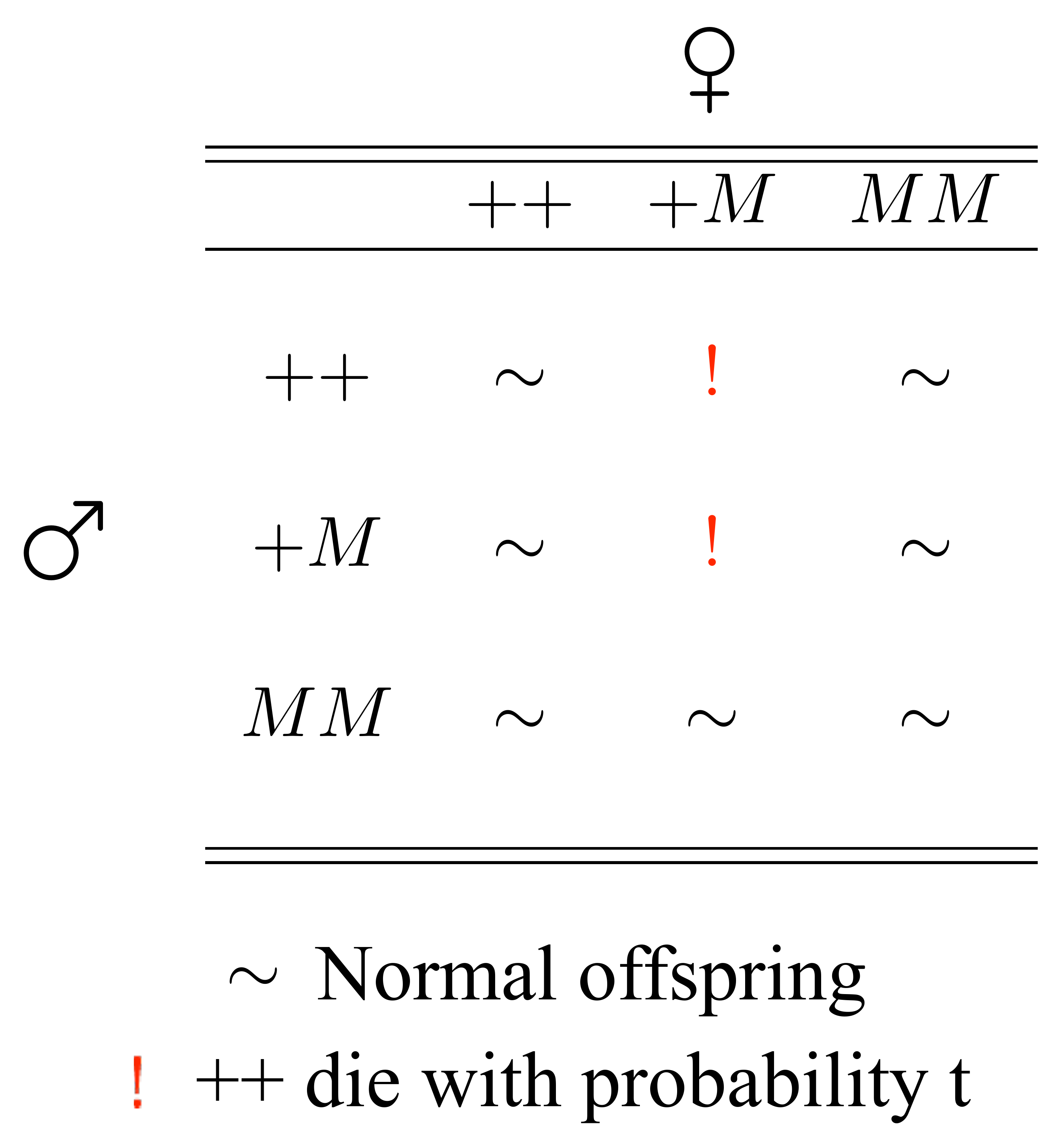}
\caption{
\label{fig:medeasystem}
The matings system of the individuals with the medea system. The wildtype homozygotes are under a lethality risk only if their maternal parent had an $M$ allele.
Thus the heterozygous females mate with wildtype homozygotes or with heterozygote males then the wildtype homozygote offspring are eliminated with probability $t$.
All other matings can produce expected number of offsprings normally.
}
\end{center}
\end{figure}

Medea is a naturally occurring selfish genetic element.
Natural \textbf{\underline{M}}aternal \underline{\textbf{e}}ffect \underline{\textbf{d}}ominant \underline{\textbf{e}}mbryonic \underline{\textbf{a}}rrest (Medea) alleles were first discovered in \textit{Tribolium} flour beetles \citep{beeman:Science:1992} and have also been reported in the mouse \citep{peters:Cell:1993,weichenhan:GR:1996}.
They derive their ability to invade populations by maternally induced lethality of wild-type offspring not inheriting a Medea allele (Fig.~\ref{fig:medeasystem}) \citep{wade:Genetics:1994}.
Thus, the wild-type homozygous offspring of the heterozygous mother die with a certain probability. 
This uniparental effect on the fitness of the offspring distorts the inheritance pattern from the usual Mendelian inheritance.
Understanding how Medea works has been a subject of interest in population genetics for long. 
It has been looked upon as a method of introducing transgenes into parasite vectors like mosquitoes. However, the genetics behind the transmission is non-Mendelian. The fitness of offspring gets affected by the genotype of its mother. 

To understand the distortion in the Medea dynamics, first we write a multiplayer game for Mendelian inheritance and then distort it.
From the point of view of an allele, it first must be present in the company of another allele in the same individual (maternal or paternal) and then when mating occurs the interaction is with two other alleles contributed by the mating partner,
which ultimately results in only two alleles being transferred to the offspring.
But first, the allele has to take into account the effects of three other copies.
In situations where the genotype of the parents matters rather than just the different genes contributed, 
there would be loss of information when considering only the alleles. 
For the alleles, we can write down a four-player game, in which the payoffs for the alleles can be denoted as given by the arrangement 
\begin{eqnarray}\label{Pmatrix1}
\begin{array}{ccccc}
\hline\hline
 & \text{\textcolor{red}{A}\textcolor{darkgreen}{A}\textcolor{blue}{A}}	&
\text{\textcolor{red}{A}\textcolor{darkgreen}{A}\textcolor{blue}{a}}	 &\highlight{
\text{\textcolor{red}{A}\textcolor{darkgreen}{a}\textcolor{blue}{a}}}	 &
\text{\textcolor{red}{a}\textcolor{darkgreen}{a}\textcolor{blue}{a}}		\\
\hline	
 \text{A} 	& a_3 &	a_2 & \highlight{a_1} & a_0	\\
 \text{a} 	& b_3 &	b_2 & \highlight{b_1} & b_0	\\
 \hline \hline
\end{array}
\end{eqnarray}
The motivation behind this particular arrangement is that even though we are looking at the frequencies of alleles, the mating occurs between two diploid individuals.
In multiplayer evolutionary game theoretic sense, this refers to a game with two strategies with four players. 
The focal player is the row player, and the combination of strategies possible for the remaining three players is given by the columns.
This arrangement helps take into account all different possibilities or variations which can be introduced by random--non-random mating, differential offspring survival or other mechanisms which can bring about a change in the allele frequencies.

Let the fitness of $AA$, $Aa$ and $aa$ be $\alpha$, $\beta$ and $\gamma$ respectively.
We focus our attention though on the frequencies of the alleles and analyse the dynamics of how they are affected by this Darwinian fitness.
The payoffs can be calculated as follows.
Consider the case of $a_1$, i.e.\ an $A$ allele interacting with the three alleles \text{\textcolor{red}{A}\textcolor{darkgreen}{a}\textcolor{blue}{a}}.

When the focal $A$ is paired with $A$ then the remaining alleles are $aa$. Mating between these two genotypes can produce only heterozygote individuals with a fitness of $\beta$. 
The other two possible combinations $Aa \times Aa$ produce one-fourth $AA$ homozygotes, one-fourth $aa$ and half of $Aa$ heterozygotes.
Hence, of the $A$ alleles in offsprings, half are in $AA$ and half in $Aa$ therefore the fitness of the $A$ allele is $( \alpha +  \beta)/2$. 
Since there are two combinatorial possibilities, we count this outcome twice.
Finally, we add up all the possible combinations, assuming random pairing and mating, and we divide the outcome by the number of possibilities of pairing the focal $A$ with $Aaa$, i.e. $3$, in order to obtain,
\begin{equation}
a_1 = \frac{\beta+( \alpha +  \beta)/2 + ( \alpha +  \beta)/2}{3} = \frac{\alpha + 2 \beta}{3}.
\end{equation}
Similarly, we can write down the rest of the payoffs, which ultimately give us the following payoff table,
\begin{eqnarray}\label{Pmatrix2}
\begin{array}{ccccc}
\hline\hline
 & \textcolor{red}{A}\textcolor{green}{A}\textcolor{blue}{A}	& \textcolor{red}{A}\textcolor{green}{A}\textcolor{blue}{a}	 &\textcolor{red}{A}\textcolor{green}{a}\textcolor{blue}{a}	 & \textcolor{red}{a}\textcolor{green}{a}\textcolor{blue}{a}		\\
 \hline
 A 	& \alpha & \frac{2 \alpha+\beta}{3}  & \frac{\alpha+2 \beta}{3} & \beta	\\
 a 	& \beta & \frac{2 \beta + \gamma}{3} & \frac{\beta+ 2 \gamma}{3} & \gamma	\\
 \hline\hline
\end{array}
\end{eqnarray}
The average payoffs simplify and reduce to,
\begin{eqnarray}
\pi_A &= \alpha x+ \beta (1-x) \nonumber \\ 
\pi_a &= \beta x + \gamma (1-x)
\end{eqnarray}
where $x$ denotes the frequency of $A$ allele.
These payoffs are often directly regarded as fitnesses in population genetics \citep{crow:book:1970}. 
In hindsight, then, we could have just made use of the payoff matrix,
\begin{equation}
\label{eq:Pmatrix}
\bordermatrix{
 & A & a \cr
A & \alpha & \beta \cr
a & \beta & \gamma \cr}
\end{equation}
which is how single-locus dynamics usually proceeds in evolutionary game theory. 
Observe that even though the Darwinian fitnesses (genotypes) are frequency independent, the allele fitnesses are not.
However, recalling our analysis, we see that this is just a special case of the four-player game which is reduced to this form.
The correspondence to two-player games is a result of the usual Mendelian inheritance seen in the example.
To display the strength of the approach illustrated here, we discuss a case of non-Mendelian inheritance, the Medea allele dynamics.

The Medea system shows a distortion in the number of offsprings generated.
Say the wild-type allele is denoted by $+$ and the Medea allele is denoted by $M$.
We assume the fitness of the wild-type $++$ to be $1$ and that of the heterozygote be $\omega$ and the Medea homozygotes have a fitness of $\nu$.
In the Medea system if an offspring is a $++$ homozygote but its mother carries an $M$ allele then the survival probability of the $++$ offspring is reduced by $t$, yielding $1-t$.

Thus the Medea element plays its role in half the cases of $++ \times +M$ and $+M \times +M$, as half the times the $+M$ will be a female and the fitness of the $++$ offspring will be $1-t$.
Applying this knowledge and the logic from above for calculating the payoffs, we can write down a separate matrix for the Medea system.

\begin{eqnarray}\label{medeamatrix}
\begin{array}{ccccc}
\hline\hline
 & \textcolor{red}{M}\textcolor{green}{M}\textcolor{blue}{M}	& \textcolor{red}{M}\textcolor{green}{M}\textcolor{blue}{+}	 &\textcolor{red}{M}\textcolor{green}{+}\textcolor{blue}{+}	 & \textcolor{red}{+}\textcolor{green}{+}\textcolor{blue}{+}		\\
 \hline
\\
 M 	& \nu & \frac{\omega + 2 \nu}{3} & \frac{2 \omega+ \nu}{3} & \omega	\\
\\
 + 	& \omega & \frac{2 \omega + 1 - t}{3}  & \frac{2 + \omega - t}{3} & 1\\
\\
 \hline\hline
\end{array}
\end{eqnarray}
Again, calculating and simplifying the payoffs give us the following relations:
\begin{eqnarray}
\pi_M &= \nu x + \omega (1-x) \nonumber \\
\pi_+ &= \omega x + (1-x t) (1-x)
\end{eqnarray}
The non-linearity is brought about in the dynamics if the $+$ allele arises naturally from considering a four-player game. 
Looking for a two-player game which reflects this scenario would make the payoff entries themselves frequency dependent.

Handling complex genetic scenarios is a regular task for multilocus population genetics theory 
\citep{feldman:Genetics:1974,karlin:Genetics:1979,christiansen:bookchapter:1988,barton:Genetics:1991}. 
However, at the same time, it can be exceedingly complex and daunting. 
We hope that bridging multilocus population genetics and multiplayer evolutionary games may help transfer simplicity of games to population genetics while taking into account the complexity of realistic genetic architectures.

\subsection*{Social sciences}

The field which has exploited the use of multiplayer game to its fullest is social sciences.
Working at the confluence of behavioural economics with cognitive scientists, psychologists and biologists, 
the social sciences provide
rich fields for experimental as well as theoretical developments.
Putting in the biological aspect in such economic reasoning especially helps in addressing the ultimate causes, the `why' questions. 
This advocates the use of evolutionary game theory in the analysis rather than classical game theory based on rationality \citep{rosas:JTB:2010}.

Addressing the evolution of cooperation has largely followed from the analysis of the famous two-player Prisoners Dilemma game \citep{axelrod:book:1984}.
However, in a social setting where interactions take place between a number of participants, the multiplayer version of the prisoners dilemma, the public goods game, has been useful \citep{hardin:Science:1968}.
An increasing number of experimental as well as theoretical investigations have brought into question the ubiquitousness of the Public goods game based on the Prisoners Dilemma. 
In turn, games such as the multiplayer stag hunt \citep{skyrms:book:2003} are thought to be appropriate in certain situations.
Instead of thinking of these games as separate instances, \citet{hauert:JTB:2006a} described them on a continuum of the so-called non-linear public goods games. 
This approach was earlier explored for a particular example of helping behaviour in \citep{eshel:AmNat:1988}. 
Understanding the social context and making use of the appropriate approximation of interactions (the game) can lead to interesting and complicated dynamics. 
\citet{archetti:JTB:2011} present a review of such non-linear fitness functions in public goods games.
Also, the concept of a threshold number of individuals required to generate a public benefit was explored in a variety of social settings \citep{bach:JTB:2006,pacheco:PRSB:2009,souza:JTB:2009}. 
Such situations, which are impossible in two-player games, have the potential to alter the evolutionary outcomes by introducing new properties to the system such as threshold values above which it is actually beneficial to cooperate.

Here, we discuss the non-linear social dilemmas as an example of multiplayer games in the social sciences.
Let us begin with the linear public goods game. 
If there are $k$ cooperators in a group of $d$ players, then they all pay a cost $c$. 
Their contribution $k c$ is multiplied  and redistributed equally amongst all the group members.
The payoff for a cooperator in this case is given by $P_C (k) = b k / d -c$ where b is the multiplied benefit ( $b = c \times \text{multiplier} $).
The defectors on the other hand do not contribute but get a share of the benefit nevertheless. 
Thus, the payoff for a defector is $P_D (k) = b k  / d $.
Note that $P_C (k)$ is defined on $k = 1, 2, \ldots, d $ while $P_D (k)$ is valid for $k = 0, 1, \ldots ,d-1 $. 
Thus, for every mixed group, defectors are better off than cooperators i.e. $P_D(k) > P_C(k)$ for $k = 1, 2, \ldots ,d-1$. 
In randomly formed groups of size $d$ the average fitness of the cooperators and defectors is then given by,
\begin{eqnarray}
f_C &=& \sum_{k=0}^{d-1} \binom{d-1}{k}x^k (1-x)^{d-1-k} P_C (k+1) \\
f_D &=& \sum_{k=0}^{d-1} \binom{d-1}{k}x^k (1-x)^{d-1-k} P_D (k)
\end{eqnarray}

where $x$ is the frequency of cooperators in the populations.
Simplifying $f_D$ using $P_D (k) = b k  / d $ we have,
\begin{eqnarray}
f_D &=&  \frac{b}{d} \sum_{k=0}^{d-1}\frac{(d-1)!}{(k-1)! (d-1-k)!}x^k (1-x)^{d-1-k}
\end{eqnarray}
Multiplying and dividing by $(d-1) x$ we get,
\begin{eqnarray}
f_D = \frac{b (d-1) x}{d} \sum_{k=0}^{d-1} \binom{d-2}{k-1} x^{k-1} (1-x)^{d-1-k} = \frac{b (d-1) x}{d}. \nonumber\\
\end{eqnarray}
In a similar way, it can be shown that $f_C = (b (x (d-1)+1)-cd)/d$.
We can actually recover the same fitnesses as of the above system by making use of a simplified two-player matrix given by
\begin{eqnarray}
\bordermatrix{ & \text{C} & \text{D} \cr
\text{C} & b-c & (b -c d)/d \cr 
\text{D} & b (d-1)/d & 0
\cr}. 
\label{reducedmat}
\end{eqnarray}
This is owing to the fact that the payoff functions are linear in the number of players, i.e. each cooperator contributed by the same amount, and hence this effect can be studied by a simplified matrix.
Instead, the dilemma could involve a non-linear payoff structure, for example each cooperator could contribute more than the previous one depicting synergistic interactions or with each cooperator the contributions decline mimicking the saturating functions \citep{hauert:JTB:2006a}.
This approach can be included in the above framework by multiplying the benefit produced by the synergy/discounting factor as $P_D (k) = b (1-\omega^k) / (d(1-\omega)) $ and $P_C (k) = P_D(k) -c $. 
For $\omega \to 1$ we recover the linear public goods game and the simplification to the two-player matrix.
For synergy ($\omega > 1$) or discounting ($\omega < 1$) we cannot simplify the average fitnesses to any meaningful two-player interpretations. 
However, \citet{pena:JTB:2014} provide a very elegant way to derive results directly from the game. 
Non-linear public goods games provide a natural way to construct intermediate cases between pure games such as the multiplayer versions of Prisoners dilemma or the stag hunt or the snowdrift game. 
While each of such pure games are amenable to simplifications to a two-player matrix the intermediates are not.

\section*{Discussion}

Developments in evolutionary game theory are possible because of a positive feedback loop between the theory and its applications \citep{broom:book:2013}.
We have just touched upon only three fields of applications of evolutionary games, but the necessity of such tools in different fields will drive the need for furthering the theory. 
For example, in ecology, the evolution of group size is still an open question which can be at least proximately addressed by evolutionary games \citep{aviles:EER:1999,pena:Evolution:2011}.
Another interesting aspect is to work on the notion of mutations.
While usually mutations are assumed to happen between existing types, completely novel mutations are hard to capture \citep{huang:NatComm:2012}.
Such mutations can also persist for long in populations under certain conditions, and it would be interesting to see how the possible stability conferred by multiplayer games interacts with the dynamic stability of multiple mutants.
Furthermore, we have limited ourselves to well-mixed populations and normal form games.
Analysis of multiplayer extensive form games is briefly mentioned in \citep{cressman:book:2003}.
The mathematics, however, becomes increasingly complicated due to the multiple game trees which are possible due to the temporally distinct actions of multiple players.
New methods \citep{kurokawa:TPB:2013} or a rediscovery of mathematical techniques such as the use of Berstein polynomials \citep{pena:JTB:2014} can make the analysis of such complicated scenarios much easier. 

Some of the finite population results discussed earlier have been extended to multiplayer games in structured populations as well \citep{van-veelen:JTB:2012b}. 
The coevolution of cooperation and multiplayer interactions has been studied albeit in a spatial context \citep{albin:Complexity:2001}.
It is not possible to take into account here the huge literature from evolutionary dynamics in structured populations
\citep{nowak:Nature:1992b,szabo:PR:2007,santos:PNAS:2006,broom:JIM:2009,broom:PRSA:2008,ohtsuki:JTB:2006b,ohtsuki:JTB:2008,van-segbroeck:PRL:2009,nowak:PTRSB:2010,van-veelen:PNAS:2012} which deserves a review of its own.  
Besides well-mixed and structured populations, equilibrium selection can also be altered by the process of random matching \citep{robson:JET:1996b,woelfing:JTB:2009} which has been extended to three-player games \citep{kaminski:BMB:2005}.

While we can directly make use of multiplayer games to survey the complex situations which nature has had to offer, how do such social interactions come about in the first place?
Exploring the evolution of multiplayer games is a new topic which may have implications for other concepts like the evolution of grouping and multicellularity. 
Complicated games can result in equally complicated or even chaotic dynamics \citep{galla:PNAS:2013}, thus putting to test the traditional concepts of evolutionary stability and other equilibrium concepts.
We still believe that studying such complications arising via multiple players is a necessity.
While the going gets tough, we quote Paul Samuelson's statement which may be valid for any growing interdisciplinary field,
\textit{``There is much territory between economics and biology that is still virgin ground. 
It will be tilled increasingly in the future. 
We should not be surprised if the first explorations are both crude and pretentious. 
Wisdom and maturity are the last settlers to arrive in pioneering communities.''} \citep{samuelson:AER:1985}.

\textbf{Acknowledgements}
We appreciate generous funding from the Max Planck Society.

\bibliographystyle{elsarticle-harv}

\bibliography{\string~/Bibtex/ET}

\end{document}